\documentclass[conference]{IEEEtran}
\IEEEoverridecommandlockouts
\usepackage[table,dvipsnames]{xcolor}
\usepackage{graphicx}
\PassOptionsToPackage{hyphens}{url}
\usepackage{hyperref}

\usepackage{booktabs}
\usepackage{fancyvrb}

\usepackage{multirow} 
\usepackage{float}
\usepackage{pifont}
\usepackage{array}
\usepackage{listings}
\usepackage{algorithm}
\usepackage[noend]{algpseudocode}
\usepackage{fancyhdr}
\usepackage{fvextra}
\usepackage[utf8]{inputenc}
\usepackage{orcidlink}
\usepackage{fontawesome5}
\usepackage{hanging}
\usepackage{pifont} 
\usepackage{threeparttable}
\usepackage{makecell}
\usepackage[T1]{fontenc}
\def\BibTeX{{\rm B\kern-.05em{\sc i\kern-.025em b}\kern-.08em
    T\kern-.1667em\lower.7ex\hbox{E}\kern-.125emX}}
\DeclareUnicodeCharacter{00A0}{ }
\DeclareUnicodeCharacter{2514}{\mbox{\kern.23em\vrule height2.2exdepth-1.8ptwidth.4pt\vrule height2.2ptdepth-1.8ptwidth.23em}}
\DeclareUnicodeCharacter{2500}{\mbox{\vrule height2.2ptdepth-1.8ptwidth.5em}}
\DeclareUnicodeCharacter{251C}{\mbox{\kern.23em
  \vrule height2.2exdepth1exwidth.4pt\vrule height2.2ptdepth-1.8ptwidth.23em}}
\makeatletter
\renewcommand{\ALG@name}{Code}
\makeatother
\definecolor{codegreen}{rgb}{0,0.6,0}
\definecolor{codegray}{rgb}{0.9,0.9,0.9}
\definecolor{makeblue}{rgb}{0.9,0.9,1.0}
\definecolor{codegray2}{rgb}{0.2,0.2,0.2}
\definecolor{codepurple}{rgb}{0.58,0,0.82}
\definecolor{backcolour}{rgb}{0.95,0.95,0.92}
\definecolor{lightgreen}{rgb}{245,255,242}

\lstdefinelanguage{Fortran2023}{
  keywords = {parallel, loop, if, in, while, do, concurrent, else,  target,enddo, end},
  comment = [l]{//},
}
\lstdefinelanguage{Compile}{
  keywords = {},
  comment = [l]{//},
}
\lstdefinestyle{mystyle}{
    backgroundcolor=\color{codegray},   
    commentstyle=\color{codegreen},
    keywordstyle=\color{codepurple},
    stringstyle=\color{codepurple},
    numberstyle=\tiny\color{codegray2},
    basicstyle=\ttfamily\footnotesize,
    breakatwhitespace=false,         
    breaklines=true,                 
    captionpos=b,                    
    keepspaces=true,                                
    showspaces=false,                
    showstringspaces=false,
    showtabs=false,                  
    tabsize=2
}
\lstdefinestyle{mystyle2}{
    backgroundcolor=\color{makeblue},   
    basicstyle=\ttfamily\footnotesize,
    breakatwhitespace=false,         
    breaklines=true,                 
    captionpos=b,                    
    keepspaces=true,                                
    showspaces=false,                
    showstringspaces=false,
    showtabs=false,                  
    tabsize=2
}
\graphicspath{{./}{figures/}}

\begin{document}
\title{Portability of Fortran's `do concurrent' on GPUs II}
\author{
\IEEEauthorblockN{Ronald M. Caplan \orcidlink{0000-0002-2633-4290}}
\IEEEauthorblockA{\textit{Predictive Science Inc.}\\
San Diego, CA USA
\\
0000-0002-2633-4290}
\and
\IEEEauthorblockN{Miko M. Stulajter \orcidlink{0000-0003-0939-1055}}
\IEEEauthorblockA{\textit{Predictive Science Inc.}\\
San Diego, CA USA
\\
0000-0003-0939-1055}
\and
\IEEEauthorblockN{Jon A. Linker \orcidlink{0000-0003-1662-3328}}
\IEEEauthorblockA{\textit{Predictive Science Inc.}\\
San Diego, CA USA
\\
0000-0003-1662-3328}
\and
\IEEEauthorblockN{Jeff Larkin \orcidlink{0000-0001-7132-3120}}
\IEEEauthorblockA{\textit{NVIDIA Corporation}\\
Santa Clara, CA USA
\\
000-0001-7132-3120}
\and
\IEEEauthorblockN{Nikolaos Tselepidis \orcidlink{0009-0008-5896-2487}}
\IEEEauthorblockA{\textit{NVIDIA Corporation}\\
Zurich, Switzerland
\\
0009-0008-5896-2487
}
\and
\IEEEauthorblockN{Harald Servat \orcidlink{0000-0002-0144-7934}}
\IEEEauthorblockA{\textit{Intel Corporation}\\
Barcelona, Catalunya ES 
\\
0000-0002-0144-7934}
\and
\IEEEauthorblockN{Shiquan Su \orcidlink{0000-0002-9546-6449}}
\IEEEauthorblockA{\textit{Intel Corporation}\\
Boulder, CO USA 
\\
0000-0002-9546-6449}
\and
\IEEEauthorblockN{Giacomo Capodaglio \orcidlink{0000-0002-6366-2656}}
\IEEEauthorblockA{\textit{Advanced Micro Devices, Inc.}\\
Austin, TX USA 
\\
0000-0002-6366-2656}
\and
\IEEEauthorblockN{Johanna Potyka \orcidlink{0000-0003-1310-4434}}
\IEEEauthorblockA{\textit{Advanced Micro Devices, Inc.}\\
Munich, Germany 
\\
0000-0003-1310-4434}
}
\maketitle

\begin{abstract}
There continues to be growing interest in using standard language constructs for parallel and accelerated HPC computing, avoiding the need for (sometimes vendor-specific) external APIs.  For Fortran applications, language features such as {\tt do concurrent} loops open the door for compilers to implement multi-threaded, GPU-accelerated, and even distributed multi-node code with only the standard language.  Here, we explore the current status of using {\tt do concurrent} for GPU-accelerated Fortran applications across three major GPU vendors (NVIDIA, AMD, and Intel).  Using a production application, we test their current capabilities, showing where the standard language alone can be used, and where augmenting the code with a directive-based API (e.g., OpenMP) is still desirable or required.  Multi-GPU tests are performed with GPU-aware MPI libraries.  We find that the three GPU vendors can now GPU-accelerate pure Fortran (zero directives), but that manual data movement directives can help with performance and compatibility.  The results show that there is rapid advancement towards making GPU-accelerated scientific HPC code performance portable using the Fortran standard language.
\end{abstract}
\begin{IEEEkeywords}
accelerated computing, do concurrent, Fortran, OpenMP, OpenACC, standard language parallelism.
\end{IEEEkeywords}

\section{Introduction}
\label{sec:intro}
The use of accelerators (such as GPUs) has become ubiquitous in high-performance computing.  For legacy codes, avoiding massive code rewrites and keeping the original code readable to domain scientists is very important.  The use of directives (such as OpenACC \cite{openacc_book} and OpenMP\textregistered\footnote{The OpenMP name is a registered trademark of the OpenMP Architecture Review Board} \cite{openmp_book}) has been a common approach towards this goal.  While less invasive, using directives still has a non-trivial learning curve to implement them correctly.  

Standard language parallelism seeks to drastically reduce or completely remove the need for external APIs or directives for parallel and accelerated computation.  Examples include C++ parallel algorithms \cite{stdparcpp,cplusplusstdparnasa,amd_hipstdpar_2024}, Fortran's {\tt do concurrent} \cite{stdpardc,inteldc} and coarrays \cite{coarrays2017,coarrays2018,coarrays2019}, and drop-in replacements for python's {\tt numpy}\footnote{\url{https://developer.nvidia.com/cupynumeric}}\cite{yadav2024automatictracingtaskbasedruntime,Chang_2026}.  While using standard parallelism does not fully remove the need for programmers to be familiar with  parallelization and vectorization concepts (such as fine-grain parallelism, structures of arrays instead of arrays of structures, etc.), no new API syntax (or vendor specific code) is required.  Additionally, since the code is within the language ISO standard, portability and longevity is built-in.

In this paper, we focus exclusively on Fortran and its {\tt do concurrent} (DC) construct for GPU-acceleration. DC loops are an alternative to {\tt do} loops that indicate the loop has no data dependencies and can therefore be computed out-of-order.  Since such a loop is likely also able to usually be computed in parallel, compilers can attempt to parallelize and off-load DC loops to GPUs (or multi-threaded CPUs). 

In our previous work from 2024~\cite{paper1}, we tested using DC with HipFT~\cite{Caplan_2025}, a finite difference code that integrates a time-dependent advection-diffusion equation on a spherical surface.  We found that it could be offloaded on all three major vendors' GPUs (NVIDIA, Intel, and AMD).  However, only NVIDIA gave optimal performance with the main code branch, while achieving good performance on Intel GPUs required small code modifications.  AMD GPUs could be targeted only with the HPE Cray compiler environment (CCE).  As described in Ref.~\cite{Caplan_2025}, CCE's DC offload feature was in development and could not yet provide expected performance.  We also found that only NVIDIA could run the code efficiently in "pure Fortran" (with zero directives), while OpenMP Target data movement directives were required for the other vendors.  Since that time, a new AMD LLVM Flang-based {\tt amdflang} compiler has been released. The bleeding edge development version of this compiler (used here) is called the advanced feature access release (AFAR)~\footnote{\url{https://github.com/amd/InfinityHub-CI/tree/main/fortran}}~\cite{amd_nextgen_fortran_2024}, which is a pre-release of the {\tt amdflang} in ROCm, that has included DC-offload for AMD GPUs. The Intel\textsuperscript{\textregistered} Fortran Compiler ({\tt ifx}), found in the Intel\textsuperscript{\textregistered} OneAPI Toolkit, has received many updates, improving performance and adding GPU-aware MPI.  The NVIDIA HPC SDK {\tt nvfortran} compiler has also received updates and improvements.  We therefore think the time is right for an update on testing the performance portability of using Fortran standard language parallelism for GPU-acceleration.

Here we use the POT3D code, a parallel finite difference preconditoned conjugate gradient (PCG) solver used in Solar physics \cite{Caplan_2021}. It is open-source\footnote{\url{https://github.com/predsci/pot3d}} and versions of it are included in the SPEChpc\textregistered 2021 \cite{spechpc,spechpc2} and SPEC CPU\textregistered 2026 \cite{speccpu,speccpu2} benchmark suites\footnote{SPEChpc\textregistered and SPEC CPU\textregistered is a trademark of the Standard Performance Evaluation Corporation}.  It has previously been shown to scale well across CPUs and NVIDIA GPUs \cite{Caplan_2021} and takes advantage of GPU-aware MPI libraries.  This allows us to test DC GPU-offload scaling across multiple GPUs and multiple multi-GPU nodes.

The paper is organized as follows:  Sec.~\ref{sec:relwork} describes related work on using Fortran standard parallelism for accelerated computing.  Sec.~\ref{sec:pot3d} describes the POT3D code and its Fortran standard parallelism implementations.  Sec.~\ref{sec:imp} shows details on how we compiled and ran the code on NVIDIA, Intel, and AMD GPUs.  Performance results for both single and multiple GPUs are shown in Sec.~\ref{sec:results}.  Sec.~\ref{sec:lib} describes linking POT3D to vendor libraries for algorithms that cannot be written in DC loops.  We conclude with summarizing the current state of the portability of Fortran standard parallelism for GPU acceleration in Sec.~\ref{sec:summary}.

\section{Related work}
\label{sec:relwork}
The use of DC for accelerated computing is a recent capability.  The first compiler to support it was NVIDIA's HPC SDK in November of 2020 \footnote{\url{https://developer.nvidia.com/blog/}\\ \url{accelerating-fortran-do-concurrent-with-gpus-and-the-nvidia-hpc-sdk/}}, followed by the Intel OneAPI Toolkit in 2022, the HPE Cray CCE in 2023, and most recently, AMD's {\tt amdflang} compiler in 2025.

Despite the recent and somewhat experimental nature of using DC for GPU offload, its use has been growing.  Examples include the Diffuse spherical smoothing tool \cite{stdpar_dc_waccpd}, the BabelStream benchmark \cite{BabelStream:Fortran}, the solar magnetohydrodynamic model MAS \cite{MAS_DC}, the relativistic magnetohydrodynamic model ECHO \cite{echodc}, the HipFT surface flux transport model \cite{Caplan_2025}, the ocean models UFDECOM-i \cite{UFDECOM-i} and A2D \cite{A2D}, and the POT3D code we use in this paper.  

All of these codes (with the exception of HipFT in our previous paper) have used DC to offload to NVIDIA GPUs only.  Here, we focus on portability, offloading to NVIDIA, Intel and AMD GPUs.

\section{POT3D}
\label{sec:pot3d}
POT3D is a finite difference based PCG solver used to compute potential field approximations of the Sun's coronal magnetic field in spherical coordinates \cite{Caplan_2021}.  The lower boundary is set to a two-dimensional surface radial magnetic field, often taken from observations.  The code is part of the CORHEL and CORHEL-CME modeling suites hosted at NASA's Community Coordinated Modeling Center, as well as in the solar wind generator tool SWiG\footnote{\url{https://github.com/predsci/swig}}.  It is open-source and available on github\footnote{\url{https://github.com/predsci/pot3d}}.

POT3D is a modern Fortran code that was originally developed for parallel computation on CPUs using MPI.  To enable GPU acceleration, OpenACC was added (as well as optional linking to the cuSPARSE library), allowing it to run on NVIDIA GPUs \cite{caplan2017mpimpiopenaccconversionlegacy}.   A version of the code was included in the SPEChpc\textregistered 2021 benchmark, where OpenMP Target directives were added (within {\tt \#IFDEF}s) for use with compilers that did not support OpenACC.  To move towards more portability, POT3D was updated by converting all computational loops into `do concurrent', removing most of the directives\footnote{\url{https://developer.nvidia.com/blog/using-fortran-standard-parallel-programming-for-gpu-acceleration}}.  The remaining directives (used for memory management and device selection) were converted from OpenACC to OpenMP Target to allow offloading to vendor GPUs whose compilers do not support OpenACC (NVIDIA also supports OpenMP Target, so no portability was lost)\footnote{The developers of POT3D prefer using OpenACC to OpenMP Target, making this required conversion not ideal, but acceptable for portability.  The conversion was helped through the use of Intel's migration tool available at  \url{https://github.com/intel/intel-application-migration-tool-for-openacc-to-openmp}}.  

An example of a DC loop within POT3D is shown in Listing~\ref{lst:dc}.
\begin{lstlisting}[language=Fortran2023, caption={Example DC loop in the POT3D code.  It computes the Laplace operator matrix-vector product for internal grid points.}, label={lst:dc},float]
do concurrent (k=2:npm1, j=2:ntm1, i=2:nrm1)
  y(i,j,k)=a(i,j,k,1)*x(i  ,j  ,k-1) &
          +a(i,j,k,2)*x(i  ,j-1,k  ) &
          +a(i,j,k,3)*x(i-1,j  ,k  ) &
          +a(i,j,k,4)*x(i  ,j  ,k  ) &
          +a(i,j,k,5)*x(i+1,j  ,k  ) &
          +a(i,j,k,6)*x(i  ,j+1,k  ) &
          +a(i,j,k,7)*x(i  ,j  ,k+1)
enddo
\end{lstlisting}
The loop is more compact then an equivalent nested {\tt do} loop, while maintaining readability and familiarity.  As mentioned, DC only indicates the loop can be run out-of-order, not necessarily in parallel.  An example of such a loop is one containing a reduction operation (such as summation, min/max, etc.). To address this, the Fortran 2023 specification added the {\tt reduce} clause to DC.  An example from POT3D is shown in Listing~\ref{lst:dcr}.
\begin{lstlisting}[language=Fortran2023, caption={Example reduction DC loop in the POT3D code.  It computes the dot product for use in the PCG solver.}, label={lst:dcr},float]
do concurrent (i=1:N) reduce(+:cgdot)
  cgdot=cgdot+x(i)*y(i)
enddo
\end{lstlisting}

A key complication that has made running standard languages on GPUs difficult is the common situation of separate memory for the CPU and GPU.  The Fortran standard does not have any notion of memory spaces, so it cannot handle data management.  As we will show, all three vendors used here have some form (virtual or otherwise) of a unified memory system where the compiler, runtime, and/or driver handles the data movement automatically.  However, it can be more efficient to manually manage the data.  In these cases, OpenACC and/or OpenMP Target have directives for this purpose.  An example is shown in Listing \ref{lst:data}.
\begin{lstlisting}[language=Fortran2023, caption={Example of using OpenMP Target for CPU-GPU data movement.}, label={lst:data},float]
! - Create array1 on GPU and copy data to GPU:
!$omp target enter data map(to:array1)
! - Use array on GPU in DC loops
!$omp target exit data map(from:array1)
! - Data is copied to CPU and array1 deleted on GPU
\end{lstlisting}

When running on multiple GPUs, even if the compiler offloads DC loops to run on a GPU, it would not know which GPU to run on (typically the runtime will pick the first GPU visible).  In order to make sure each MPI rank is running on the correct GPU, we can use OpenACC and/or OpenMP Target directives in combination with the MPI local shared rank as shown in Listing \ref{lst:device}.
\begin{lstlisting}[language=Fortran2023, caption={Directives used to set the GPU device number in POT3D assuming 1 GPU per MPI rank, where {\tt iprocsh} is the MPI shared rank.}, label={lst:device},float]
!$ call omp_set_default_device (iprocsh)
!$acc set device_num(iprocsh)
\end{lstlisting}
To accomplish the same objective without these directives, the system can be set to expose only a single GPU for each MPI rank.   A bash script to do this is included with POT3D and shown in Listing \ref{lst:launch}.
\begin{lstlisting}[language=bash, mathescape=false, caption={Launch script to run a code with 1 GPU per local MPI rank.}, label={lst:launch},float]
#!/bin/bash
#===> Uncomment correct LOCAL_RANK definition
#===> and device selector for your system:
#  => SLURM:
LOCAL_RANK=${SLURM_LOCALID}
#  => PBS / Torque:
# LOCAL_RANK=${PBS_VNODENUM}
#  => OpenMPI:
# LOCAL_RANK=${OMPI_COMM_WORLD_LOCAL_RANK}
#  => MPICH / Intel MPI / Hydra:
# LOCAL_RANK=${PMI_LOCAL_RANK}
#  => Intel OneAPI MPI >=25:
# LOCAL_RANK=${MPI_LOCALRANKID}
#===> Make local MPI rank see the GPU to use:
#  => NVIDIA:
export CUDA_VISIBLE_DEVICES=${LOCAL_RANK}
#  => AMD ROCm:
#export ROCR_VISIBLE_DEVICES=${LOCAL_RANK}
#  => AMD HIP:
#export HIP_VISIBLE_DEVICES=${LOCAL_RANK}
#  => LLVM:
#export OMP_DEFAULT_DEVICE=${LOCAL_RANK}
#  => Intel:  Either use oneAPI / SYCL ...
#export ONEAPI_DEVICE_SELECTOR=level_zero:${LOCAL_RANK}
#  => ... or Level Zero:
#export ZE_AFFINITY_MASK=${LOCAL_RANK}
#export I_MPI_OFFLOAD_PIN=0 (Intel)
#===> Launch:
exec "$@"
\end{lstlisting}

Unlike the HipFT code from our previous work, POT3D uses GPU-aware MPI for multi-GPU communication (and for periodic boundary conditions even on a single GPU).  GPU-aware MPI allows data residing on the GPU to be passed directly to the MPI library.  In the absence of a unified memory system, this can be accomplished by using OpenMP's {\tt use\_device\_addr} directive to send the GPU arrays to the MPI library as shown in Listing \ref{lst:gpumpi}.
\begin{lstlisting}[language=Fortran2023, caption={Example use of GPU-aware MPI with OpenMP Target.}, label={lst:gpumpi},float]
!$omp target data use_device_addr(array1)
call MPI_Allreduce (MPI_IN_PLACE,array1,N, &
                    MPI_REAL8, MPI_SUM,    &
                    MPI_COMM_WORLD,ierr)
!$omp end target data
\end{lstlisting}

To test the portability of using DC in a pure Fortran code, we have created a branch of POT3D called {\tt stdpar} that has zero directives (requiring a unified memory system).  We compare this branch with one that adds OpenMP Target directives for CPU-GPU manual data movement and device selection, denoted as {\tt stdpar\_ompdata}\footnote{The "main" branch of POT3D is the same as {\tt stdpar\_ompdata}, but uses classic {\tt do} loops with OpenMP Target directives for all reductions.  This is because older versions ($<15$) of the GNU GCC {\tt gfortran} compiler that are ubiquitous on many systems do not recognize the Fortran 2023 "reduce" clause on DC loops.}.

\section{Implementation}
\label{sec:imp}
Here we describe how we build and run POT3D on GPUs from three major vendor (NVIDIA, Intel, and AMD).  For each vendor, we use their respective 1st-party compilers as they are often the most reliable/optimal for new features.  All of the vendors' compilers shown here are freely available.  We note that there are other compilers that have begun to support DC for GPU-acceleration, including HPE's Cray Compiler Environment\footnote{\url{https://cpe.ext.hpe.com}} and mainline LLVM\footnote{\url{https://flang.llvm.org}}.

\subsection{NVIDIA}
\label{sec:nvidia}

The way to compile Fortran DC codes for acceleration on NVIDIA GPUs using the NVIDIA HPC SDK compiler suite has remained stable since the detailed description in our previous paper \cite{paper1}.  We recap the basics here.   Offloading DC to GPUs is activated with the compiler flag {\tt -stdpar=gpu}.  Specification of the GPU type/model and/or the way to handle GPU memory is set with the {\tt -gpu=} flag.   When compiling on the same system the GPU resides on, we use the {\tt -gpu=ccnative} to target the resident GPU; otherwise, we specify the GPU explicitly (e.g., {\tt cc80} for an A100).  If using OpenMP Target directives for device selection and data movement, the additional flags {\tt -mp=gpu} and {\tt -gpu=mem:separate} are needed.  Due to the OpenMP and OpenACC runtimes not "seeing" each others' device selection, both OpenMP and OpenACC device selection directives are used (as shown in Listing~\ref{lst:device}), and the {\tt -acc=gpu} flag is added. 

NVIDIA has a robust unified memory management feature that allows one to not need directives to manually manage the data between GPU and CPU.  This unified memory is active by default (or for verbosity, can be set with {\tt -gpu=mem:unified}), and works with both unified devices (e.g., Vera-Rubin superchip, GB200) and discrete GPUs (e.g., B300, RTX Pro 6000).  For discrete GPUs, heterogeneous memory management (HMM) must be enabled on the system, otherwise the older {\tt -gpu=mem:managed} memory option can often be used.

For using GPU-aware MPI, there exist many MPI libraries that support NVIDIA GPUs\footnote{\url{https://developer.nvidia.com/mpi-solutions-gpus}}.  The easiest way to get started is by using the NVIDIA HPC SDK's included HPC-X \cite{hpcx} based on the OpenMPI library, pre-configured for GPU-aware MPI.  With the included HPC-X, no special environment variables are needed to activate the GPU-aware features.

The unique compiler flags and environment variables needed for each branch of POT3D used for NVIDIA GPUs for the runs in this paper are shown in Table~\ref{tab:comp}.

\subsection{INTEL}
\label{sec:intel}
The basic compiler flags to accelerate DC loops on Intel GPUs remains stable from our previous work \cite{paper1}, but we summarize them here for completeness.  Offloading DC to GPUs is activated with the compiler flag {\tt -fopenmp-target-do-concurrent}, which requires having OpenMP enabled with the flag {\tt -fiopenmp}.  Specification of the target GPU is done with the {\tt -fopenmp-targets=} flag, where we set it to {\tt spir64} which covers all compatible GPUs at the time of this writing.  In order to avoid CPU-GPU copies on every DC loop (i.e., tell the compiler that the data management is taken care of either through OpenMP target data directives or unified memory), it is critical to set the flag {\tt -fopenmp-do-concurrent-} {\tt maptype-modifier=present}.

While the OneAPI compiler allows for DC offload to Intel GPUs, the use of OpenMP Target data movement has been required for efficient performance for most codes (e.g., the {\tt stdpar\_ompdata} branch of POT3D).  Recently, Intel has implemented a unified memory system called shared system USM that is designed to automatically manage the GPU and CPU memories~\cite{servat2026unifiedsharedmemoryopenmp}, allowing pure Fortran codes to run on Intel GPUs.  It is included in the Intel runtime 26.18\footnote{\url{https://github.com/intel/compute-runtime}} and in the Xe driver within the Linux kernel ($\ge 6.19$) for GPUs supporting page faulting (starting with Xe2).  In OneAPI ($\ge 2026.1$), it is activated with the compiler flag {\tt -fopenmp-force-usm}.

For using GPU-aware MPI, the most mature implementation is Intel's own MPI library (IMPI), but there is support in other libraries as well (e.g., HPE's Cray MPICH, etc.).  The IMPI library comes with the Intel OneAPI toolkit, so it is the easiest way to get started.  In order to activate the GPU-aware MPI features, the environment variable {\tt I\_MPI\_OFFLOAD} must be set to 1 or 2, while additional variables may be needed for efficient performance or based on system configuration.  The extra variables we used are shown in  Table~\ref{tab:comp}, where we also list the unique compiler flags needed for each branch of POT3D for Intel GPUs.

\subsection{AMD}
\label{sec:amd}
In our previous work \cite{paper1}, we were able to successfully run a Fortran DC code with OpenMP Target data movement directives on AMD GPUs with the HPE CCE compiler.  However, as the feature was very new, there was not yet enough performance for reasonable use.  Since then, CCE has had several updates, but for this work, we did not have access to a system with the latest CCE compiler to test.  
Within the same time frame, an implementation of DC offload to AMD GPUs has been added to AMD's LLVM-based compiler called {\tt amdflang}.  While {\tt amdflang} comes with AMD ROCm installations (both the production releases\footnote{\url{https://rocm.docs.amd.com/en/latest}} and the preview releases\footnote{\url{https://rocm.docs.amd.com/en/7.13.0-preview}}), the versions that support DC GPU offload are the stand-alone  AOMP\footnote{\url{https://github.com/ROCm/aomp/tree/aomp-dev}} and the AFAR\footnote{\url{https://github.com/amd/InfinityHub-CI/blob/main/fortran}} for the pre-release {\tt amdflang} compiler package integrated in a pre-release ROCm. While DC offload features are being added to all versions of {\tt amdflang} (and {\tt flang}), for this paper, we use the AFAR package.  This {\tt amdflang} compiler supports both manual data movement with OpenMP directives, as well as the ability to use unified shared memory for AMD GPUs that support it \footnote{At present, only data center AMD GPUs and APUs support unified shared memory (e.g., MI250X, MI300A), while consumer and professional graphics cards do not.} 

Offloading DC to GPUs is activated with the compiler flag {\tt-fdo-concurrent-to-openmp=device}, which requires having OpenMP enabled with the flag {\tt -fopenmp}.  Specification of the target GPU is done with the {\tt --offload-arch=} flag (e.g., {\tt gfx942} for the MI300A).

To use unified memory (either direct for APUs like the MI300A, or virtually with HMM for GPUs like the MI250X), no special compiler flags are required.  Instead, it is activated by setting the environment variable {\tt HSA\_XNACK=1}.  The unified memory allows for efficient performance without any data movement directives.

For GPU-aware MPI on AMD GPUs, there is support from several MPI libraries.  The AFAR/ROCm compiler package does not include a pre-configured GPU-aware MPI library. However, AMD has made available helpful scripts that build MPI libraries with full ROCm AMD GPU-aware support\footnote{\url{https://github.com/amd/HPCTrainingDock/tree/main/comm/scripts}}.  Here, we build OpenMPI with the AMD-provided script {\tt openmpi\_setup.sh} modified to build against HPE's slingshot libfabric library.  With this build, no special environment variables are required to activate GPU-aware MPI.

The compiler flags and environment variables needed for each branch of POT3D used for AMD GPUs for the runs in this paper are shown in Table~\ref{tab:comp}.
\begin{table*}[htbp]
\centering
\begin{threeparttable}
    \centering
    \begin{tabular}{|l|l|l|}
        \hline
        \rowcolor{lightgreen}
        \textbf{Code Branch} &
        \textbf{Compiler Flags} &
        \textbf{Environment Variables Needed\tnote{a}} \\
        \hline
        \multicolumn{3}{|c|}{\color{ForestGreen}\textbf{NVIDIA ({\tt nvfortran})}}\\
        \hline
        \makecell[tl]{
            OpenMP Target Data\\
            (\texttt{stdpar\_ompdata})
        }
        &
        \makecell[tl]{
            \texttt{-stdpar=gpu} \\
            \texttt{-mp=gpu} \\
            \texttt{-acc=gpu} \\
            \texttt{-gpu=ccnative,mem:separate}
        }
        &
        \makecell[tl]{
            No non-default required.
        } \\            
        \hline
        \makecell[tl]{
            Pure Fortran \\
            (\texttt{stdpar})
        }
        &
        \makecell[tl]{
            \texttt{-stdpar=gpu} \\
            \texttt{-gpu=ccnative,mem:unified}\tnote{e}
        } &
        Same as above. \\
        \hline
        \multicolumn{3}{|c|}{\color{Blue}\textbf{INTEL ({\tt ifx})}}\\
        \hline
        \makecell[tl]{
            OpenMP Target Data\\
            (\texttt{stdpar\_ompdata})
        }
        &
        \makecell[tl]{
              \texttt{-fiopenmp} \\
              \texttt{-fopenmp-target-do-concurrent} \\
              \texttt{-fopenmp-targets=spir64} \\
              \texttt{-fopenmp-do-concurrent-}\\
              \texttt{ maptype-modifier=present}
        }
        &
        \makecell[tl]{
              \texttt{I\_MPI\_OFFLOAD=1} \\
              \texttt{ZE\_FLAT\_DEVICE\_HIERARCHY=COMPOSITE}\tnote{b} \\
              \texttt{I\_MPI\_OFFLOAD\_SYMMETRIC=1} \\
              \texttt{I\_MPI\_OFFLOAD\_TOPOLIB=none} \\
              \texttt{I\_MPI\_OFFLOAD\_DOMAIN\_SIZE=1} \\
              \texttt{LIBOMPTARGET\_DEVICES=SUBDEVICE}\tnote{b} \\
              \texttt{I\_MPI\_OFFLOAD\_IPC=0}\tnote{c} \\
              \texttt{I\_MPI\_FABRICS=ofi:ofi}\tnote{c} \\
        } \\
        \hline
        \makecell[tl]{
            Pure Fortran \\
            (\texttt{stdpar})
        }
        &
        \makecell[tl]{
             \texttt{-fiopenmp} \\
             \texttt{-fopenmp-target-do-concurrent} \\
             \texttt{-fopenmp-targets=spir64} \\
             \texttt{-fopenmp-do-concurrent-}\\
             \texttt{ maptype-modifier=present} \\
             \texttt{-fopenmp-force-usm}
        }
        &
        \makecell[l]{Same as above} \\
        \hline
        \multicolumn{3}{|c|}{\color{Red}\textbf{AMD ({\tt amdflang})}}\\
        \hline
        \makecell[tl]{
            OpenMP Target Data\\
            (\texttt{stdpar\_ompdata})
        }
        &
        \makecell[tl]{
              \texttt{-fopenmp} \\
              \texttt{-fdo-concurrent-to-openmp=device} \\
              \texttt{--offload-arch=gfx942}
        }
        &
        \makecell[tl]{
              \texttt{ulimit -s unlimited} \\
              \texttt{HSA\_XNACK=0} \\
              \texttt{UCX\_MEMTYPE\_CACHE=n}\tnote{d} \\
              \texttt{HSA\_ENABLE\_SDMA=0}\tnote{d} \\
              \texttt{UCX\_RNDV\_SCHEME=get\_zcopy}\tnote{d} \\
        } \\
        \hline
        \makecell[tl]{
            Pure Fortran \\
            (\texttt{stdpar})
        }
        &
        Same as above
        &
        \makecell[tl]{
              Same as above, but with: \\
              \texttt{HSA\_XNACK=1}
        } \\
        \hline
    \end{tabular}

    \begin{tablenotes}
    \item[a] These are the non-default environment variables we needed for the runs based on the systems and MPI libraries we ran on. Other systems may require a different set of variables or none at all.
    \item[b] These should only be set for multi-tile GPUs like the MAX 1550.
    \item[c] These environment variable were set due to a security update on Stampede3 that disabled IPC.
    \item[d] These environment variables were required for GPU-aware MPI to work on consumer AMD GPUs.
    \item[e] Alternatively, one can use \texttt{-gpu=mem:managed} which may be needed on older systems that do not have HMM support.
    \end{tablenotes}
\end{threeparttable}
    \vspace{8pt}
    \caption{Compiler flags and environment variables used to run both branches of POT3D on NVIDIA, Intel, and AMD GPUs.}
    \label{tab:comp}
\end{table*}
\addtocounter{footnote}{0}

\section{Results}
\label{sec:results}

POT3D includes several example and benchmark runs.  Here, we use two benchmark runs. The first, called {\tt bench\_tiny}, uses $\sim 73.2$ million cells and has identical inputs to the SPEChpc\textregistered 2021-Tny {\tt 528.pot3d\_t} benchmark run\footnote{\url{https://www.spec.org/hpc2021/docs/benchmarks/528.pot3d_t.html}}.  The second run has identical inputs to the SPEChpc\textregistered 2021-Sml {\tt 628.pot3d\_s} benchmark run\footnote{\url{https://www.spec.org/hpc2021/docs/benchmarks/628.pot3d_s.html}} and is called {\tt isc2023} in the POT3D repository.  It uses $\sim 300$ million cells and was used for the 2023 ISC student cluster competition \cite{isc2023scc}.  The smaller {\tt bench\_tiny} benchmark is used to test single GPUs and single CPU nodes to ensure the implementations are performing on the order expected across vendors and hardware.  The larger {\tt isc2023} benchmark is used to test multi-GPU and multiple multi-node scaling.

For all runs, we do not intend to show the most optimized performance possible.  Instead, we look for functionality, correctness, and "reasonable" performance of DC across the various systems and hardware without any vendor specific tuning applied.  Available systems have a variety of compilers, drivers, network setup, configurations, etc.  Additionally, some DC features are still designated as "experimental" within some compilers. Therefore, the results should not be used to compare or evaluate the general performance of each vendors' hardware and software. 

\subsection{Single GPU}
\label{sec:singlegpu}
To test the DC implementations across vendors, we use the POT3D {\tt bench\_tiny} benchmark.  POT3D's performance is extremely memory bandwidth bound \cite{spechpc2}.  Therefore, to check that the DC implementations on GPUs are performing at reasonable levels, we run the benchmark on four CPU systems spanning a wide range of memory bandwidths to create a linear performance trend.  We then run on the GPUs and see if they perform at expected levels based on their memory bandwidths.  If so, we classify the run as `acceptable performance.' 

\begin{figure*}[htbp]
    \centering
    \includegraphics[width=0.75\linewidth]{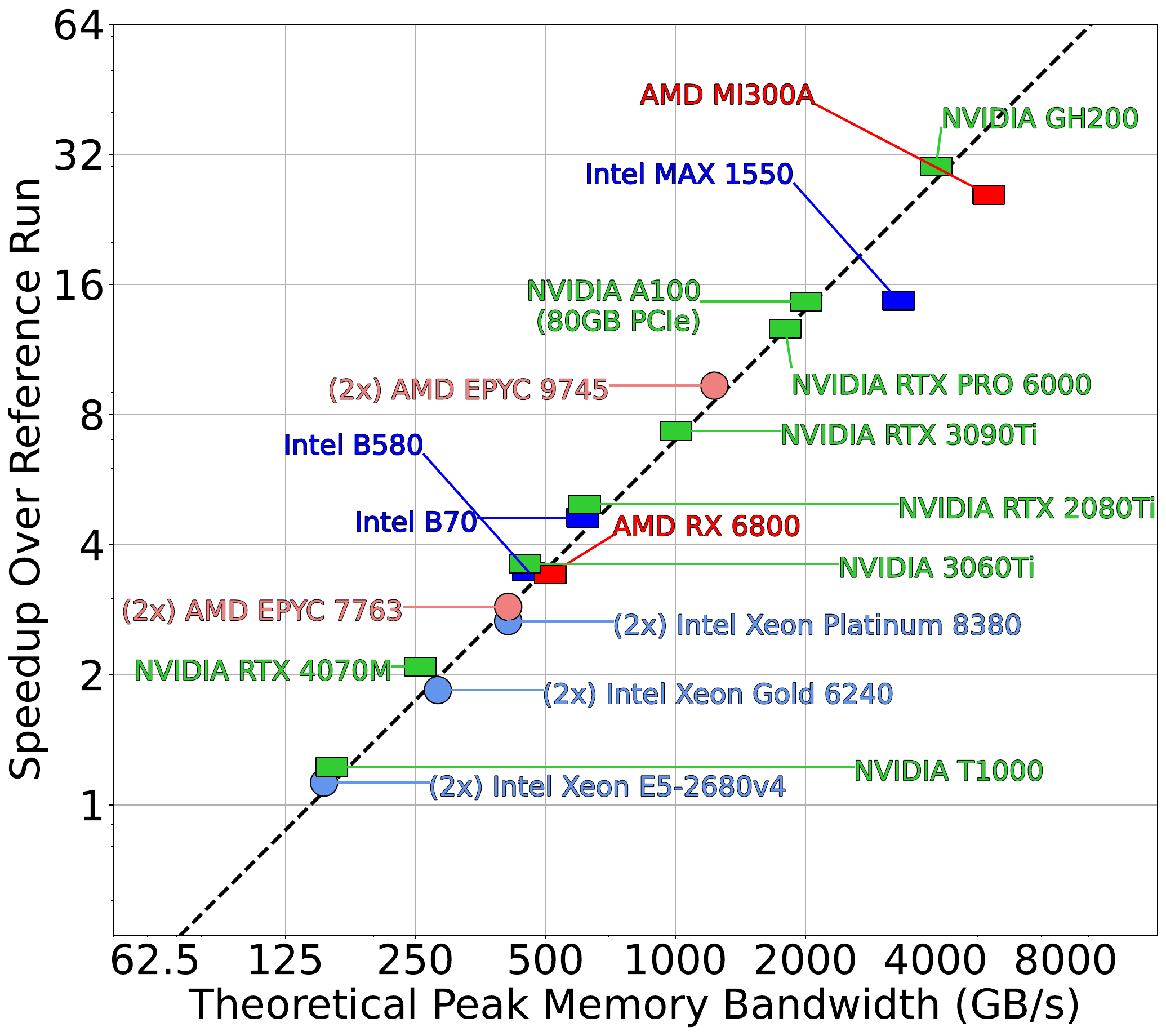}
    \caption{Performance of the POT3D {\tt bench\_tiny} benchmark (same inputs as SPEChpc\textregistered 2021-Tny {\tt 528.pot3d\_t}) for single GPUs (rectangles) and single CPU nodes (circles). The results are plotted as speedup over a reference run time (higher is better) as a function of theoretical peak memory bandwidth.  CPU runs use the pure Fortran {\tt stdpar} branch, while GPU runs use the OpenMP Target data directive {\tt stdpar\_ompdata} branch.  We see that the GPU DC implementations are overall performing reasonably.  The Intel MAX 1550 used 2 MPI ranks (one per tile), while the AMD MI300A used 6 MPI ranks (one per XCD in CPX mode {--} see Sec.~\ref{sec:multigpu}).
    \label{fig:tiny_speedup_vs_memband}}
\end{figure*}

In Fig.~\ref{fig:tiny_speedup_vs_memband} we show the wall clock speedup of the {\tt bench\_tiny} benchmark compared to a reference run time (2125 seconds) as a function of theoretical peak memory bandwidth.\footnote{It should be kept in mind that extremely memory bandwidth bound codes like POT3D have vastly different performance profiles compared to FLOP-bound, mixed, or AI/tensor/low-precision workloads across different architectures.  Therefore, a compute hardware that appears faster than another in this test, may be slower in other workloads, and vise-versa.}   For all CPU runs, we use the {\tt stdpar} branch of POT3D (pure Fortran with zero directives), while for all GPU results, we use the {\tt stdpar\_ompdata} branch. Note that to obtain results on a variety of compute hardware, the compiler and driver version varies between the results.
We see that generally, all three vendors have GPUs that perform as expected based on the memory bound performance trend.  The Intel 1550 GPU (which is no longer available) has been reported to have an issue achieving near-peak memory bandwidth \cite{10820802}, causing it to drift from the expected performance trend.   The MI300A also does not fall on the expected performance trend line.  It did achieved near the expected performance for the compute sections of the code (68 seconds), but the internal MPI communication with 6 ranks in CPX mode (see Sec.~\ref{sec:multigpu}) caused extra overhead time (15 seconds). 

\begin{figure}[htbp]
    \centering
    \includegraphics[width=\linewidth]{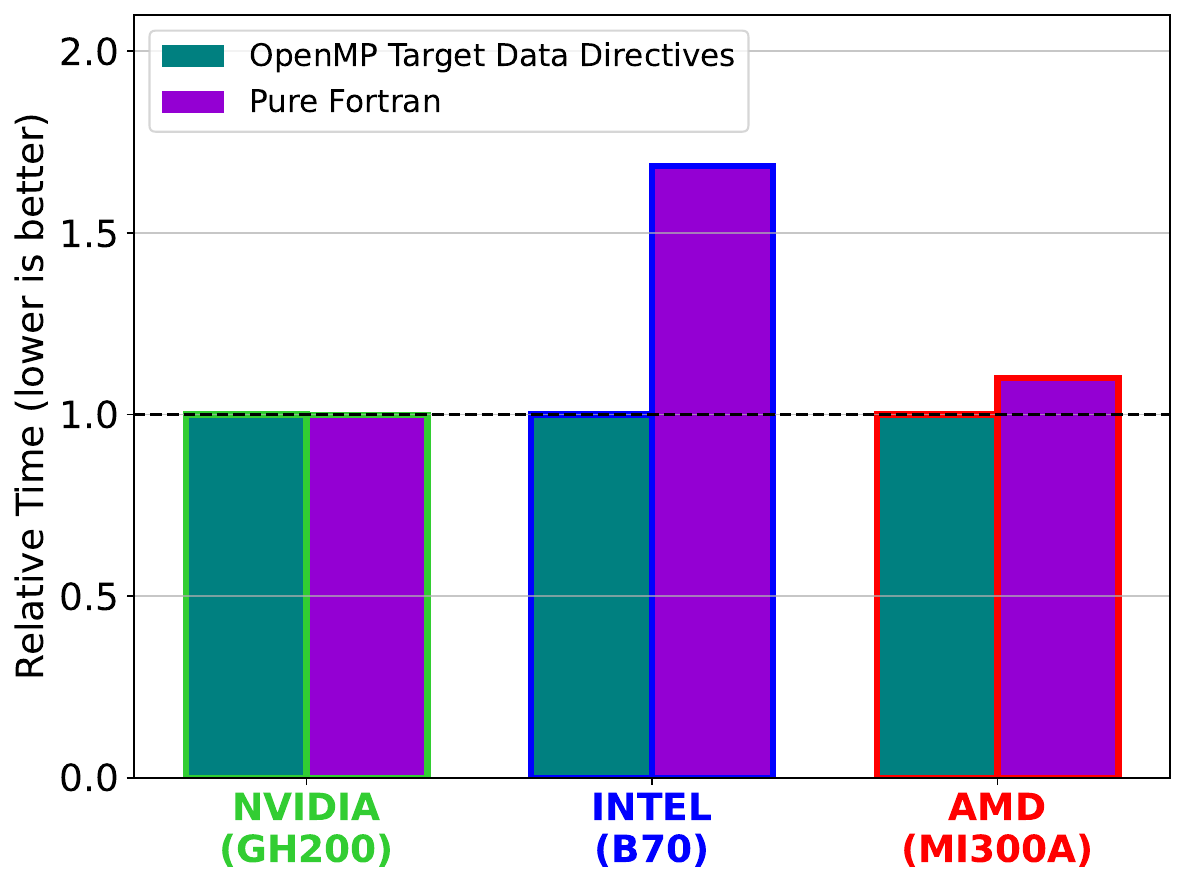}
    \caption{Relative speed (lower is better) of the POT3D {\tt bench\_tiny} benchmark (same inputs as SPEChpc\textregistered 2021-Tny {\tt 528.pot3d\_t}) running with the {\tt stdpar\_ompdata} (left) and {\tt stdpar} (right) branches on a single GPU.  The {\tt stdpar} runs (pure Fortran) use their vendors' respective unified memory capabilities.}
    \label{fig:omp_vs_std}
\end{figure}

While the DC GPU results for {\tt stdpar\_ompdata} show that all three vendors can now offload DC loops in Fortran to GPUs, they used OpenMP Target directives for data movement.  To test the ideal case (pure Fortran with no directives), we run the same test on a GPU from each vendor, but using the {\tt stdpar} pure Fortran branch (utilizing unified memory).  The results are shown in Fig.~\ref{fig:omp_vs_std}.
We see that all three vendors are able to run the pure Fortran code on GPUs.  For NVIDIA, the performance of the pure Fortran code is slightly \emph{faster} than  using OpenMP Target directives.  For AMD, the performance between the branches is very close, while for Intel, the pure Fortran is over $1.5\times$ slower than the manual data movement version.  However, the ability to use unified memory for pure Fortran codes on Intel GPUs is a cutting-edge feature of the compiler, driver, and runtime, so the performance gap is expected to be reduced with future development. 

We conclude this section by emphasizing that GPU offloading of DC works across data center, professional, and consumer GPUs\footnote{For AMD with AFAR, the pure Fortran {\tt stdpar} code (unified memory) only currently works on data center GPUs that support {\tt HSA\_XNACK}.}, allowing one to develop standard parallel Fortran codes and perform small-to-medium runs on a GPU-equipped workstation, and perform large runs on large GPU supercomputers/data centers.

\subsection{Multiple GPUs}
\label{sec:multigpu}
Using multiple GPUs can be accomplished by using MPI to spread the work out, and running the local portion of the work of an MPI rank on its own GPU (oversubscribing can also be done but is beyond the scope of this paper).  An important difference between the multi-GPU HipFT code used in our previous work and POT3D, is that POT3D uses MPI communication on arrays that are expected to reside on GPU memory.   This requires either manually paging data back and forth to the CPU around MPI calls, or (ideally) using a GPU-aware MPI implementation.  When using OpenMP Target data directives (as in the {\tt stdpar\_ompdata} branch) GPU arrays can be sent to the GPU-aware MPI calls through the use of the {\tt use\_device\_addr()} directive as we showed in Listing~\ref{lst:gpumpi}.  When using pure Fortran with a unified memory system (e.g., the {\tt stdpar} branch), the combination of the compiler, runtime, drivers, and network setup have to automatically handle the data for  the MPI calls.  This is hidden from the user, so it is not easy to know if the GPU arrays end up being paged to the CPU, or if the GPU data is sent across the network directly without profiling the code.

Here, we run the POT3D {\tt isc2023} benchmark across multiple GPUs and multiple GPU nodes.  The strong scaling performance of multi-GPU codes like POT3D is heavily dependent on the system configuration, both hardware (e.g., networking) and software (e.g., drivers, runtime, and compiler versions, as well as environment setup and MPI library installation and availability).  Therefore, in the present context, we are more interested in functionality and "reasonable" scaling, than trying to achieve perfect scaling.  We are focused on how using pure Fortran with unified memory can affect the scaling of the code, not on the original scaling of the  specific case and setup.

\begin{figure}[htbp]
    \centering
    \includegraphics[height=1.12\linewidth]{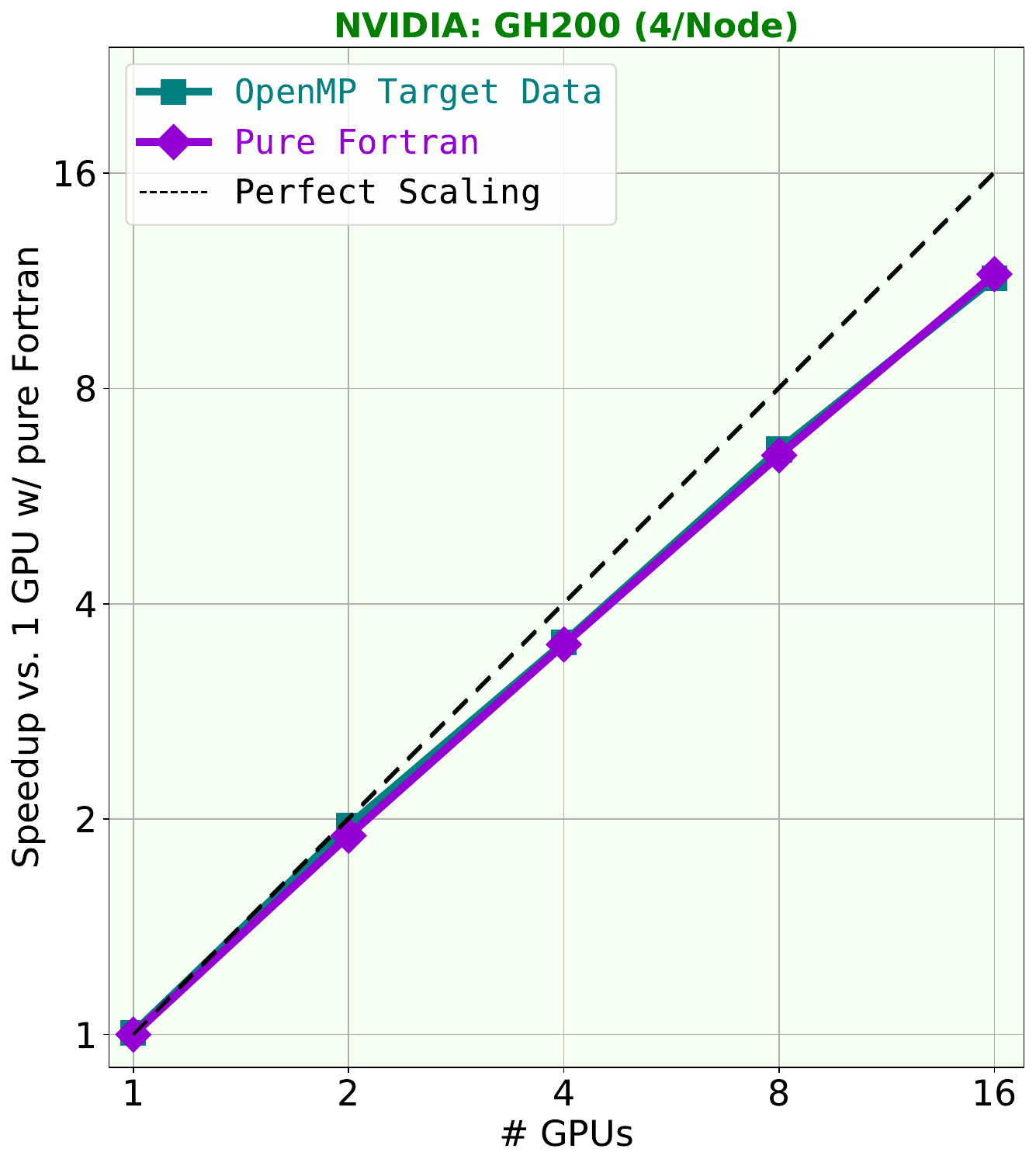}
    \caption{Scaling performance of the {\tt isc2023} benchmark run for multiple NVIDIA GH200 GPUs. The runs were performed on the JUPITER system at the Jülich Supercomputing Centre.  It used the NVIDIA HPC SDK Fortran compiler {\tt nvfortran} 25.9 and the OpenMPI library 5.0.8, with the NVIDIA driver 595.71.05 and CUDA runtime 13.2.}
    \label{fig:scaling_nvidia}
\end{figure}

In Fig.~\ref{fig:scaling_nvidia}, we show the scaling of the {\tt isc2023} benchmark from 1 to 16 GH200 GPUs (with nodes that have 4 GPUs per node).
We see that the code scales well out to 16 GPUs (4 nodes).  The pure Fortran code with unified memory results in nearly identical run times compared to the manual data management OpenMP Target branch.

\begin{figure*}[htbp]
    \centering
    \includegraphics[height=0.56\linewidth]{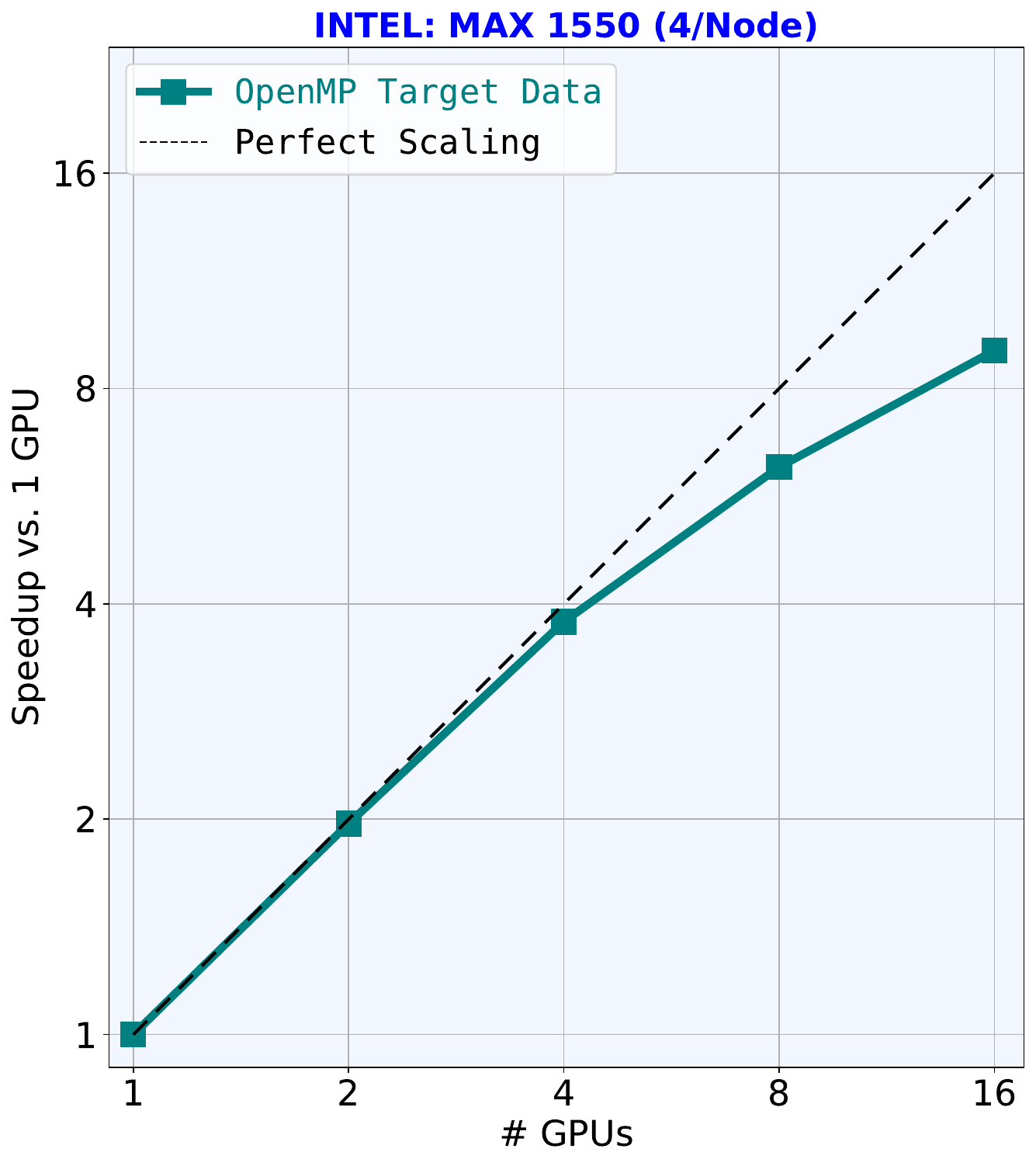}
    \includegraphics[height=0.56\linewidth]{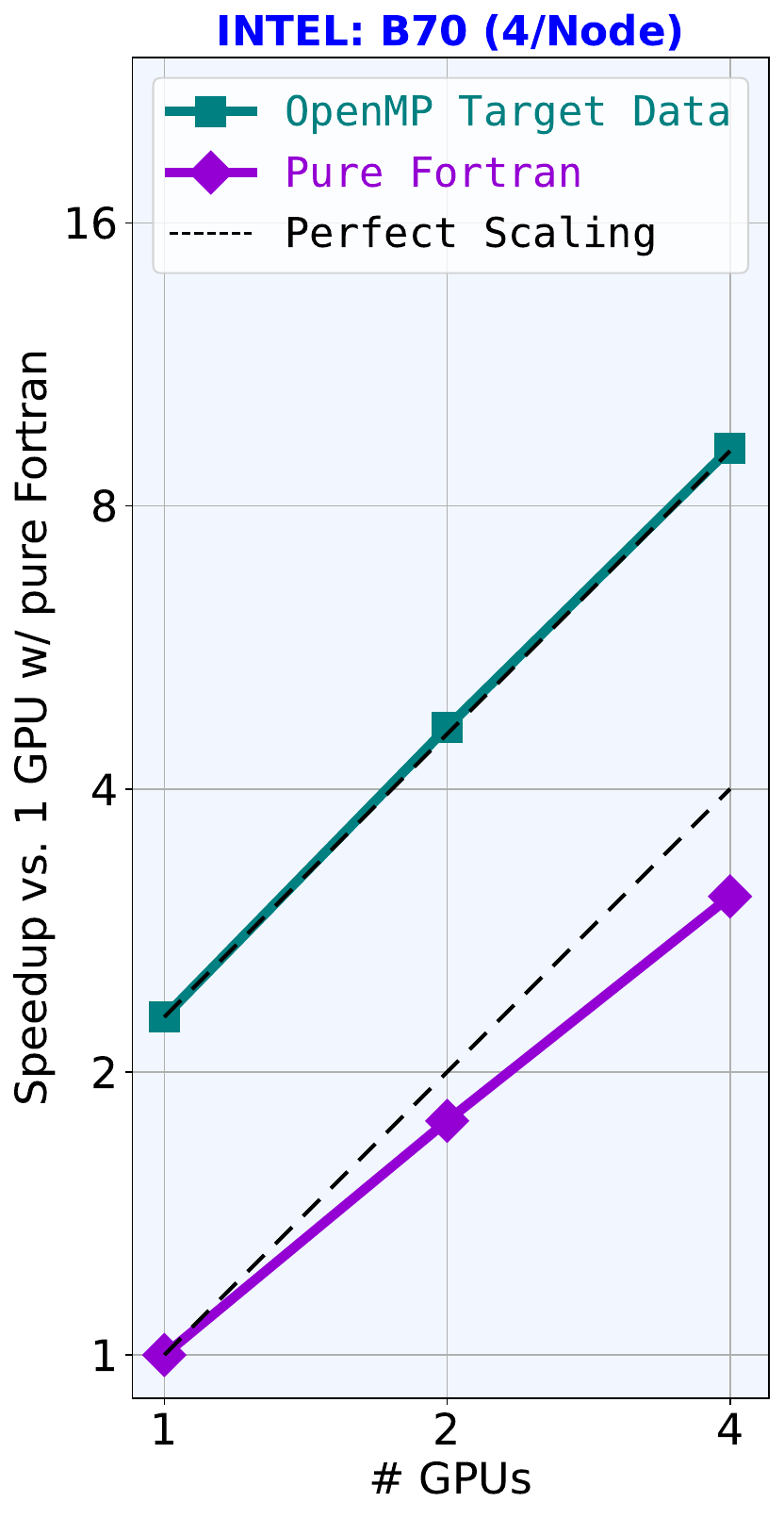}
    \caption{Scaling performance of the {\tt isc2023} benchmark run for multiple Intel MAX 1550 (left) and B70 (right) GPUs. The MAX 1550 runs were performed on the Stampede3 system at TACC.  It used the OneAPI {\tt ifx} compiler 2026.0.0 and the IMPI library 21.18, with the I915 driver 25.2.43\_PSB\_250224.50. The B70 results were performed on an internal Intel test system with the OneAPI toolkit 2026.1, the Linux kernel 7.1.0-rc6 and its Xe driver, and the 26.18.38380.1 compute runtime.}
    \label{fig:scaling_intel}
\end{figure*}

In Fig.~\ref{fig:scaling_intel}, we show the scaling for two models of Intel GPUs.  The Intel\textsuperscript{\textregistered} Data Center GPU Max 1550 is the most recent Intel data center GPU that is deployed on a supercomputer, allowing us to test multi-node runs.  It consists of two compute tiles, which can be configured such that the GPU is seen by the runtime as a single GPU, or where each tile is seen as a separate GPU.  In our runs in this paper, we use the latter configuration, using 1 MPI rank per tile.  The MAX 1550 GPU does not support the new unified memory features of the Intel driver.  We therefore also test the scaling of POT3D on Intel\textsuperscript{\textregistered} Arc\textsuperscript{\texttrademark} Pro B70 GPUs which do support unified memory.  However, we are limited to testing the scaling of the B70s on a single node/workstation (as they are professional GPUs, not data center GPUs).
We see that the MAX 1550 scales well across multiple nodes, and that the B70 scales perfectly within a single workstation/node for the {\tt stdpar\_ompdata} branch.  The pure Fortran unified memory runs on the B70s scale well across multiple GPUs, but have a consistent significant slowdown compared to the manual memory management runs, as was the case in Fig.~\ref{fig:omp_vs_std}.  The ability to use pure Fortran with unified memory is an extremely new feature for Intel GPUs, and this performance is expected to improve with further driver/runtime/compiler development.

\begin{figure}[htbp]
    \centering
    \includegraphics[height=1.12\linewidth]{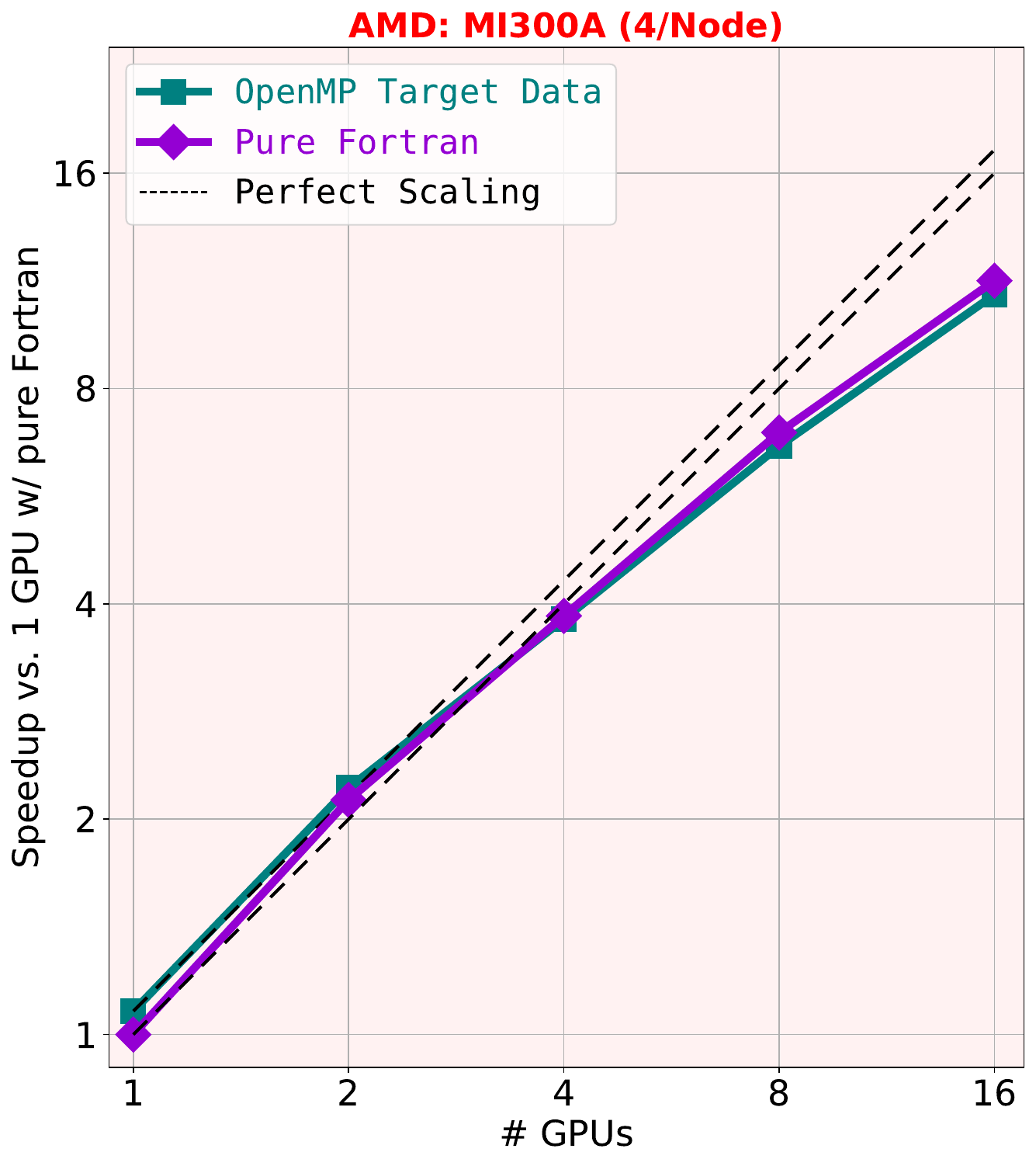}
    \caption{Scaling performance of thew {\tt isc2023} benchmark run for multiple AMD MI300A APUs. The runs were performed on the AAC7 internal AMD system with the {\tt amdflang} compiler from the AFAR "therock" package 23.2.1 including ROCm v7.13.0, HSA runtime 1.21, and OpenMPI 5.0.10.}
    \label{fig:scaling_amd}
\end{figure}

In Fig.~\ref{fig:scaling_amd}, we show the scaling results for the AMD MI300A up to 16 APUs with 4 APUs per node.  The MI300A consists of 6 accelerator complex dies (XCDs) and can be configured to run in "SPX" mode where the entire GPU is treated by the runtime as a single device, or in "CPX" mode where each XCD is treated as separate devices (there also exists a TPX mode where the GPU is treated as 3 GPUs) \cite{amd_mi300_partition_modes_2025}.  For the result in Fig.~\ref{fig:tiny_speedup_vs_memband}, we used CPX mode with 6 MPI ranks.  Here, we use "SPX" mode (1 MPI rank per GPU).
We see good scaling overall, both intra- and inter-node,  similar to the results from other vendors.  The unified memory pure Fortran runs are nearly the same performance as the manual memory runs, and in some cases are slightly faster.

\section{External libraries}
\label{sec:lib}
Another method to avoid the use of external APIs or directives is calling pre-made libraries optimized for GPU performance for each vendor for supported algorithms.  While these libraries are often coded in vendor-specific low-level languages, from the point of view of the domain scientist, it can be a simple subroutine call within the Fortran code.  In this section, we briefly illustrate using such libraries for the optional ILU0 preconditioner in POT3D.

As mentioned in Sec.~\ref{sec:pot3d}, POT3D uses a preconditioned conjugate gradient solver.  It features two preconditioners:  1) point-jacobi/diagonal scaling (PC1) and 2) non-overlapping domain decomposition ILU0 (PC2).  PC2 reduces the number of solver iterations much more than PC1, but costs more per iteration to apply.  PC1 is easily vectorized, and therefore simple to implement with DC.  All results in this work until here have therefore used PC1.  The standard/simple algorithms for implementing PC2 (both the incomplete LU factorization to set up the PC, and the required triangular solves to implement it per iteration) are inherently sequential and cannot be directly implemented with DC.  Parallel algorithms (including those efficient on GPUs) have been developed over the years but can be complicated to implement in the standard languages, often requiring the use of lower level vendor-specific languages.  Due to the potential performance advantage of using ILU0, all three GPU vendors tested here have external libraries that can be called to implement PC2.  NVIDIA has the cuSPARSE library (released in 2010), Intel has the OneAPI MKL SYCL library (released in 2021), and AMD has the rocSPARSE/hipSPARSE  library (released in 2016).  There are also open source projects that could run PC2 on all three vendors, such as GINKO \cite{ginkgo-toms-2022}.

POT3D has had PC2 available on NVIDIA GPUs using cuSPARSE for many years.  Here, we have completed preliminary (non-optimized) implementations of PC2 on GPUs on Intel and AMD GPUS using their respective sparse libraries.  It is beyond the current scope to describe the details of these implementations, but overall, it involves writing a Fortran interface to a small C code file that contains the calls to the libraries\footnote{There exist direct Fortran bindings to some libraries, but we prefer minimizing the footprint in the main Fortran code}.  The added code to these calls is minimized and are put within {\tt \#IFDEFS}.

\begin{figure}
    \centering
    \includegraphics[width=0.45\linewidth]{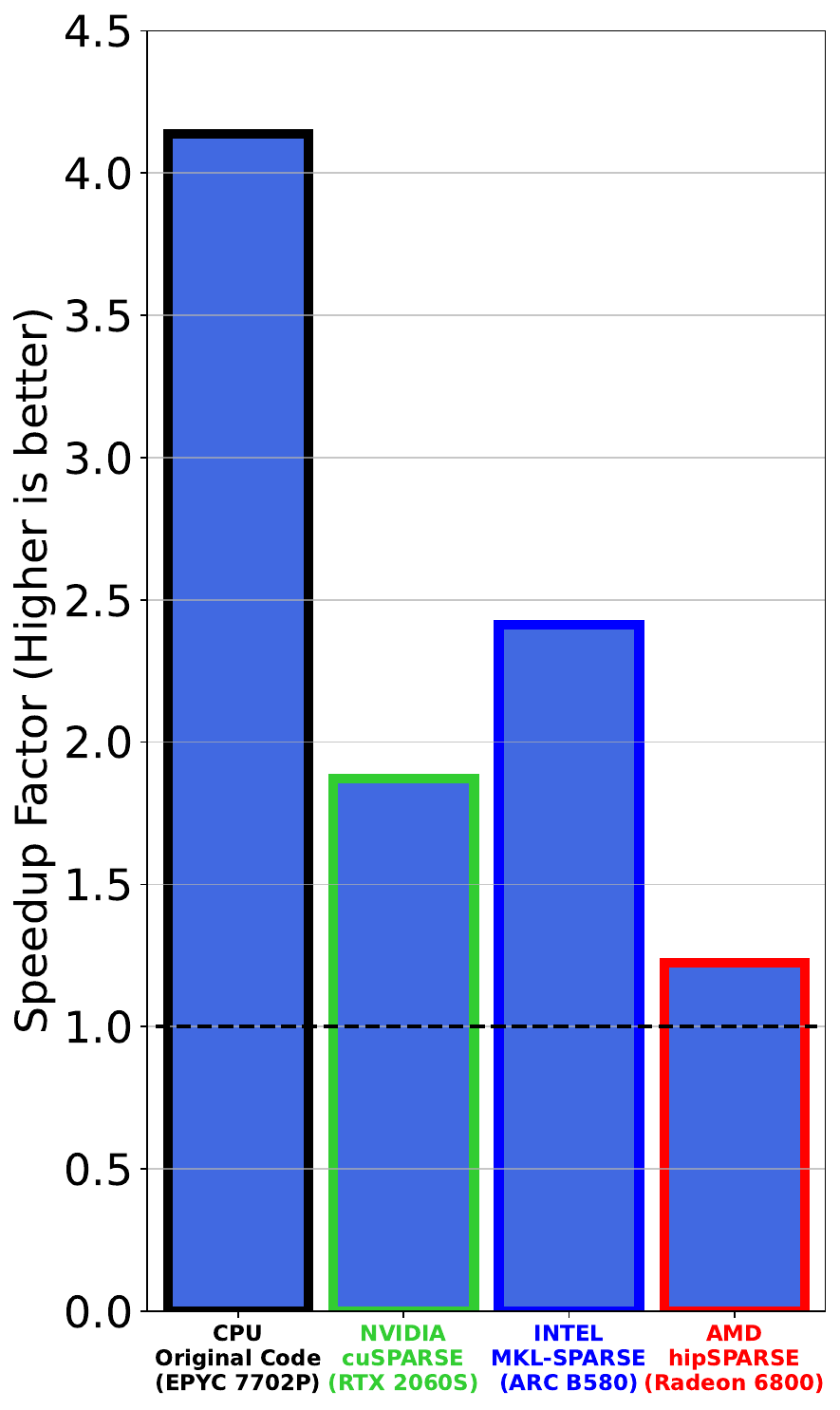}
    \includegraphics[width=0.45\linewidth]{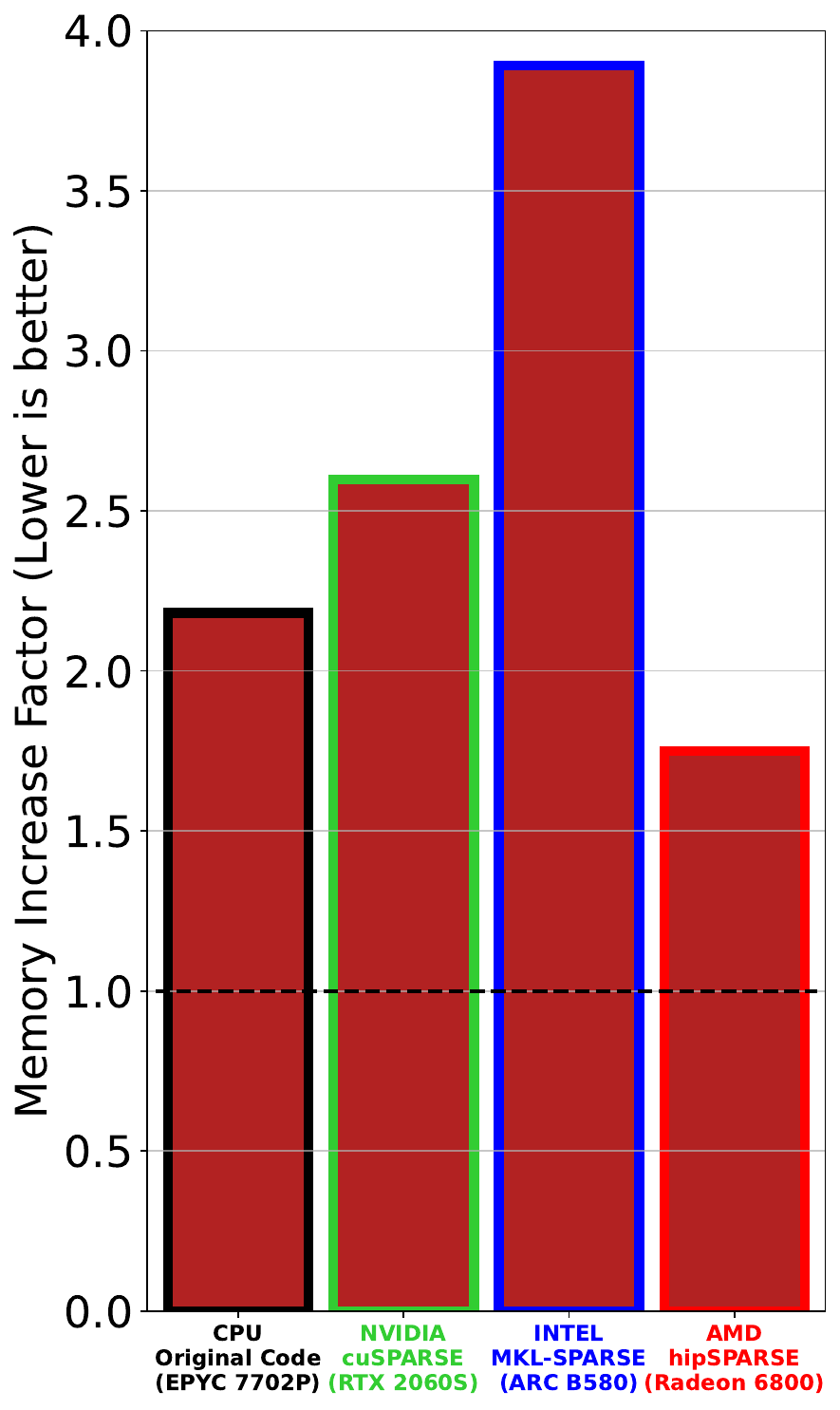}    
    \caption{Relative speedup (left) and memory increase factor (right) using the PC2 (ILU0) preconditoner compared to the PC1 (diagonal scaling) preconditioner on the same hardware.  Results used the {\tt stdpar\_ompdata} branch with the {\tt open\_field} testsuite run.  PC2 is implemented with each GPU vendors' sparse matrix library.}
    \label{fig:pc1_vs_pc2}
\end{figure}

In Fig.~\ref{fig:pc1_vs_pc2} we show the results of using the {\tt stdpar\_ompdata} branch of POT3D on the {\tt open\_field} testsuite run using PC1 and PC2 on a CPU and across the three vendors' GPUs\footnote{We note that, currently, the Intel OneAPI MKL does not have a GPU-accelerated ILU0 factorization routine, so that is done on the CPU at the start of the code and transferred to the GPU for use with the GPU-enabled triangular solvers.}.
We see that the PC2 runs on GPUS are up to two times faster than PC1, while on CPU, PC2 is 4 times faster.  This is consistent with our previous experience \cite{caplan2022banded} and is a result of the difficulty of making fully parallel versions of the ILU0 factorization and triangular solvers for narrow-banded matrices on GPUs.  We also note that PC2 uses more memory than PC1, and on GPUs, can use significantly more (up to 4x in this case).  This increase in memory usage reduces the maximum problem size that can be run on the GPUs.

Although the speedup of PC2 over PC1 on the GPUs is not as great as that on CPUs, it can still be significantly faster ($\sim 2\times$), and therefore the ability to call these libraries from within the Fortran DC code is very useful.  We note that not all vendors currently support using these library calls from a pure Fortran code using unified memory, and that the speedup results shown here can be higher or lower based on problem size and hardware support.

\section{Summary and outlook}
\label{sec:summary}
In this work we have demonstrated running a production Fortran application (POT3D) on GPUs, without requiring the use of language extensions or external APIs (e.g., pure ISO Fortran).  We were able to run across three major GPU vendors (NVIDIA, Intel, and AMD) using each vendors' freely available compiler. The code uses the Fortran language's {\tt do concurrent} (DC) loop construct, which allows the compilers to parallelize the loop on both multi-threaded CPUs an GPUs.  Through the use of GPU-aware MPI, we were also able to run the code across multiple GPUs.  Management of data between GPU and CPU memory was handled by each vendors' unified/auto-paging memory features.  

To evaluate the performance of the pure Fortran GPU-offload runs, we compared them to runs using a version of the code that includes manual OpenMP Target data management directives.  We found that on NVIDIA and AMD GPUs, the two versions had very similar performance, even on multiple GPUs.  On Intel GPUs, the pure Fortran code ran over 50\% slower, however the Intel unified memory feature is extremely new and is expected to improve with subsequent compiler updates. 

While we continue to use the OpenMP Target data directive branch of the code in production until the performance of the pure Fortran code is more robust across systems and vendors, the results shown here show massive improvements in the portability and performance of ISO Fortran for accelerated computing.   We expect to be able to not require any directives, language extensions, or higher-level abstraction languages for our codes in the coming years.  This will greatly increase the productivity of domain scientists who can continue to develop in their familiar language, with only small changes in coding style (DC vs. {\tt do} loops) and design (making sure loop iterations are independent).

We also have demonstrated the use of Fortran calls to external vendor libraries to allow the use of advanced algorithms that cannot be expressed in DC loops that will run well on GPUs.  This allows a pure Fortran code to take advantage of libraries written in lower-level/vendor-specific languages, without having the developer learn such languages.

While we did not have access to every GPU model, every network stack, every software/compiler stack and environment, the variety of vendor and hardware we were successful in running Fortran with DC on GPUs shows that DC has grown in portability and performance since our last investigation \cite{paper1}, especially in the addition of the {\tt amdflang} compiler allowing offloading to AMD GPUs, and Intel's new unified memory features.

Future compiler development could take advantage of ISO Fortran features that are not commonly known or used by some Fortran developers.  Specifically, the use of  "coarrays," which can allow Fortran codes to run across multiple compute nodes without the explicit use of MPI or other message-passing libraries.  Combined with DC loops, coarrays could allow running on many GPUs across multiple compute nodes without MPI.  This has been referred to as the `dream' of Fortran becoming a fully portable, intrinsically parallel language \cite{Rouson2024}.

Our next step in using standard language parallelism in Fortran will be running our Magnetohydrodynamic simulation code MAS across the three vendors' GPUs.  MAS uses DC loops and currently runs on NVIDIA GPUs \cite{MAS_DC}.  It is a more complicated (and much larger) code base than POT3D (e.g., derived types, function calls within DC loops, etc.), making it an ideal next step to test the portability of Fortran DC on GPUs.

\section{Artifact Availability Statement}

\newcommand*\ttvar[1]{\texttt{\expandafter\dottvar\detokenize{#1}\relax}}
\newcommand*\dottvar[1]{\ifx\relax#1\else
\expandafter\ifx\string/#1\string/\allowbreak\else#1\fi
\expandafter\dottvar\fi}

\section*{Artifact Identification}
We have continued investigating the portability of using Fortran's `do concurrent` (DC) for accelerated computing by testing its use across GPU vendors using the solar physics potential field code POT3D.  POT3D is written using DC loops as the primary method of GPU offloading.  Two branches are used, one which includes OpenMP target directives for manual data movement between the CPU and GPU and device selection, and one that has zero directives (pure Fortran).  We successfully compiled and ran production-level tests on NVIDIA, Intel, and AMD GPUs using NVIDIA's HPC SDK's {\tt nvfortran}, Intel's OneAPI HPC Toolkit's {\tt ifx}, and AMD's AFAR {\tt amdflang} compilers respectively.  

Tags/releases of the github repository of POT3D (\url{https://github.com/predsci/pot3d}) were made for the versions used in this paper, specifically the \url{https://github.com/predsci/POT3D/releases/tag/v4.7.0_stdpar_ompdata} and  \url{https://github.com/predsci/POT3D/releases/tag/v4.7.0_stdpar} releases.   The test cases are located in the {\tt benchmarks/bench\_tiny} and {\tt benchmarks/isc2023} folders of the POT3D github repository, and correspond to the SPEChpc\textregistered 2021 "tiny" and "small" runs respectively.

Given access to the computational hardware and driver/runtime/library stack, the repository with its release tag can be used to directly reproduce results shown in the paper by building with the described compiler flags, along with any/all system-specific compiler flags and environment variables that may be needed.  The POT3D code outputs the time a run took in a file called {\tt timing.out}, as well as writing it to the terminal.

\section*{Reproducibility of Experiments}
The workflow to generate the results in this paper is as follows:  First, checkout the two branch releases of POT3D.  The release for {\tt  stdpar\_ompdata} is located at \url{https://github.com/predsci/POT3D/releases/tag/v4.7.0_stdpar_ompdata} and and the release for {\tt stdpar} is located at  \url{https://github.com/predsci/POT3D/releases/tag/v4.7.0_stdpar}.  Next, build the codes according to the build instructions (including having an HDF5 library installed with the same compiler used for POT3D), using the included sample build scripts to make build scripts with the compiler flags described in the paper for the specific system/hardware being used.  Then, copy the input files from the {\tt benchmarks/bench\_tiny}, {\tt benchmarks/isc2023}, and/or {\tt testsuite/open\_field/input} folders into a new folder, and run the code (located in {\tt bin/pot3d}) with the MPI launcher used on your system.  The code should be run with 1 MPI rank per CPU core for CPU runs, and 1 MPI rank per GPU or sub-GPU device (e.g. Intel tile or AMD XCD) for GPU runs.

The execution time will depend on the hardware and software platforms.  If running on similar hardware as described in the paper, the run times should be very close to those in shown in the figures (some variation due to speed of the file system and network may be encountered).  The results show the portability of Fortran's DC for GPU acceleration, with similar timings for NVIDIA, Intel, and AMD GPUs of comparable capabilities (for memory bound algorithms like those in POT3D).

\section{Acknowledgments}
Work at Predictive Science Inc.\ was supported by the NASA LWS Program (grants 80NSSC22K0893 and 80NSSC24K1108).  It also utilized the Cabeus system at
NASA’s HECC through NASA grant’s 80NSSC20K0192’s
NAS request SMD-24-72380598, as well as the Stampede3 system at TACC through allocation TG-MCA03S014 from the Advanced Cyberinfrastructure Coordination Ecosystem: Services \& Support (ACCESS) program, which is supported by U.S. National Science Foundation grants \#2138259, \#2138286, \#2138307, \#2137603, and \#2138296.

System access to JUPITER was provided as part of the JUPITER Research and Early Access Program, JUREAP. JUPITER is supported by the EuroHPC JU and GCS through funding by the European Commission, the German Federal Ministry of Research, Technology and Space, and the Ministry of Culture and Science of the State of North Rhine-Westphalia.

We would like to thank Bob Robey, Michael Klemm, Kareem Ergawy, and the co-authors from AMD for their invaluable help with the pre-release AFAR {\tt amdflang} compiler and granting access to the AAC7 and AAC6 MI300A test systems at AMD. We also thank AMD's Hakob Arzumanyan for his valuable help setting the correct environment to enable GPU-aware MPI on consumer AMD GPUs.


\bibliographystyle{IEEEtran}
\bibliography{ref.bib}

\end{document}